\documentclass[prl,showpacs,superscriptaddress,amsfonts,amsmath,floatfix,twocolumn]{revtex4-1}
\usepackage{graphicx}
\usepackage{hyperref} 
\usepackage{verbatim} 
\usepackage{epstopdf}
\usepackage[normalem]{ulem}
\usepackage{color}
\usepackage{dsfont}

\newcommand{\affA}{Center for Integrated Quantum Science and Technology, Institute for Complex Quantum Systems, Universit\"at Ulm, D-89069 Ulm, Germany}
\newcommand{\affB}{Instituto de F{\'i}sica da UFRGS, Av. Bento Gon{\c c}alves 9500, Porto Alegre, RS, Brazil}

\begin{document}
\title{Contact and Static Structure Factor for Bosonic and Fermionic Mixtures}

\author{Rafael E. Barfknecht}
\email{rafael.barfknecht@ufrgs.br}
\affiliation{\affB}

\author{Ioannis Brouzos}
\email{ioannis.brouzos@uni-ulm.de}
\affiliation{\affA}

\author{Angela Foerster}
\email{angela@if.ufrgs.br}
\affiliation{\affB}

\date{\today}
\begin{abstract}
\noindent We study measurable quantities of bosonic and fermionic mixtures on a one-dimensional ring.
These few-body ensembles consist of majority atoms obeying certain statistics (Fermi or Bose) and an impurity atom in a different hyperfine state. 
The repulsive interactions between  majority-impurity and majority-majority are varied from  weak to strong. 
We show that the majority-impurity repulsion is mainly responsible for the loss of coherence in the strongly interacting regime.
The momentum distribution follows the $\mathcal{C}/p^4$ universal behaviour for the high momentum tail, but the contact $\mathcal{C}$ is strongly dependent on the strength of the majority-impurity and in a different way on the  majority-majority interactions. 
The static structure factor of the majority atoms exposes a low-momentum peak for strong majority-impurity repulsion,  which is  attributed to an effective attraction not expected for purely repulsive forces.

\end{abstract}

\pacs{67.85.-d, 02.30.Ik, 03.75.Hh}

\maketitle

\section{Introduction}

Non-local measurable quantities of cold atom systems play a major role in the experimental probe of their properties. 
From the momentum distributions in the first realizations of Bose-Einstein condensates \cite{ketterle,cornell} and the time-of flight images of phase transitions in optical lattices \cite{mott}, to the probing of the Tonks-Girardeau (TG) gas (an infinitely repulsive one-dimensional bosonic ensemble with fermionic properties \cite{paredes,weissTG}), experimental techniques have given rise to an unprecedented degree of control, manipulation and measurement of atomic systems \cite{papp,thalhammer,haller}. 
For one-dimensional systems, theoretical research has developed analytical and numerical methods to study properties of those quantities, like the universal $\mathcal{C}/p^4$ asymptotic behaviour of the momentum distribution \cite{minguzzi,girardeau,calabrese}, which is governed by the contact $\mathcal{C}$. This is a notion that captures all universal properties of such systems even close to phase transitions \cite{zhou}, and  has been recently measured in ultracold gases \cite{vale1,jin,vale2}.
The structure factor, another important concept originated in solid-state physics to probe crystalline lattices, has recently been studied \cite{astra,panfil,stringari} and observed for 1D Bose gases via Bragg spectroscopy \cite{fabbri}.

More recent advances in experiments deal with mixtures of Bose-Bose, Bose-Fermi and Fermi-Fermi cold gases \cite{myatt,catani,mccarron,pagano}. 
It became also possible to realize few-body ensembles of bosons \cite{fewchinese,bourgain} and fermions in different hyperfine states \cite{selim}, and measure the effects of an impurity by increasing the number of fermions one by one \cite{fewtomany}. 
For composite mixtures of bosons and fermions several local quantities have been studied and have exhibited various phases not present in pure ensembles \cite{adhi,astra_mixtures,zinner}, but some advances are still to be done on the side of non-local correlators \cite{zwerger}.

In this work we investigate, using a Jastrow-type ansatz, some measurable non-local quantities for a few-body ensemble of bosons or fermions in a one-dimensional ring in the presence of an impurity of the same mass but in a different hyperfine state. In these systems, repulsive interactions between impurity-majority or - for the bosonic case - majority-majority pairs can be tuned via Feshbach resonances. For the integrable system constituted by an impurity in a Fermi sea, the ground state wave functions and energies have been exactly obtained in \cite{mcguire}, and results for correlations are discussed in \cite{recher}. For a system of bosons, if all interaction strengths are equal, the Lieb-Liniger integrable model can be applied \cite{lieb}. However, when the interaction strengths differ and the system becomes non-integrable \cite{yiannis-angela,sutherland,lamacraft}, we show that all measurable non-local correlation quantities exhibit a behaviour substantially different from that of the integrable one. 
In particular, the reduced one-body density matrix of the majority atoms, a correlation function that shows the degree of coherence in its off-diagonal terms, is strongly dependent on the impurity-majority coupling strength. At the same time, the high momentum tail of the momentum distribution obeys a $\mathcal{C}/p^4$ universal behaviour for all cases studied here. The numerical values for the contact $\mathcal{C}$ depend not only on the impurity-majority and majority-majority couplings, but also on the nature of the atom being considered.
For different interactions, the static structure factor deviates from the integrable case, above and below the phononic behaviour of the Tonks-Girardeau gas; most surprisingly, a pronounced peak arises for strong majority-impurity repulsion, which is an indicator of a quasi-crystalline structure and of effective attractive correlations in a purely repulsively interacting ensemble. 
The results obtained here for the correlations via a simple Jastrow ansatz are compared with the exact solutions in the corresponding cases where those solutions exist, showing in general very good agreement. 

\section{System Hamiltonian and Ansatz}

We consider a system of 3 atoms on a one-dimensional ring of length $L$ with contact interactions and periodic boundary conditions. 
The Hamiltonian reads
\begin{equation} 
\label{hamiltonian}
H=-\frac{1}{2}\sum_{j=1}^{3}\frac{\partial^2}{\partial x_{j}^2}+\sum_{i<j} g_{ij}\delta(x_{i}-x_{j})
\end{equation}
where the lengths are in units of $L$ and energies in units of $\hbar^2/mL^2$. 
The interaction strength $g_{ij}$ may be different for each pair of atoms $i,j$ and is controllable via Feshbach or confinement induced resonances since $g=g_{1D}/(\hbar^2/mL)$ with $g_{1D}= \frac{2\hbar^2 a_{3D}}{m a^2_{\perp}} \left(1-\frac{|\zeta(1/2)| a_{3D}}{\sqrt{2} a_{\perp}}\right)^{-1}$, where $m$ is the atom mass, $\zeta$ is the Riemann zeta function, $a_{3D}$ is the 3D s-wave scattering length and $a_{\perp}$ the length of the transversal confinement \cite{cigar}. The latest is assumed to be very small such that the system is effectively 1D.
The composite system where the only difference between the atoms is the coupling strengths $g_{ij}$ with respect to each other can be realized experimentally by using atoms of the same species in different hyperfine states.

\begin{figure}
\includegraphics[scale=0.45]{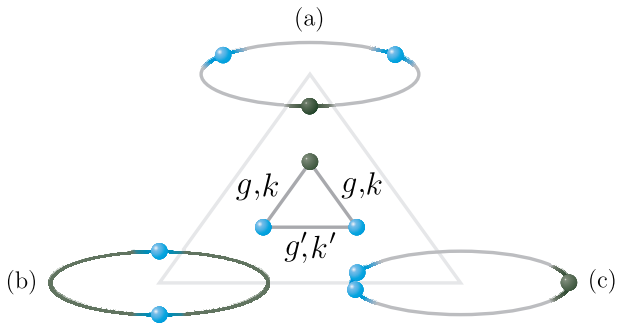} 

\caption{(Color online) Schematic depiction of the physical system for representative cases: $(a)$ strongly repulsive integrable case, in which the separation between the three atoms is at maximum. (b) The interaction between the majority pair is strong, but the impurity-majority interaction is weak; therefore, the impurity atom appears ``delocalised'' across the ring. (c) The impurity strongly repels the majority pair but the majority-majority interaction is weak, so these atoms tend to bunch together on the other side of the ring.}
\label{fig1}
\end{figure}

As shown in Fig.~\ref{fig1} we consider 3 atoms on a ring, with two of them being in the same hyperfine state and obeying fermionic or bosonic statistics, and the third (which we call the impurity) in a different hyperfine state. Therefore we have two important parameters, the coupling strength $g$ between impurity and majority atoms and the $g'$ for the bosonic case for the interaction between the majority atoms. In the case where the majority atoms are fermions, the s-wave scattering is forbidden due to the statistics, and the only parameter that remains is the impurity-majority coupling strength $g$. 

\subsection{Underlying Physics}
In Fig.~\ref{fig1} we schematically depict the extreme cases that indicate the basic underlying physical scenarios that appear in this system. Fig.~\ref{fig1} (a) represents the integrable case where $g=g'$ and the atoms are strongly repulsive, so they tend to maximize the distance between them. In Fig.~\ref{fig1} (b), only the majority atoms repel strongly each other (or obey the Pauli exclusion principle if they are fermions), and therefore they tend to localize on diametric positions on the ring. The impurity is delocalised all over as a single particle since the interaction $g$ with the others is vanishing. On the other hand, in Fig.~\ref{fig1} (c) the interaction between the bosonic majority atoms $g'$ is vanishing or very weak, and the impurity strongly repels both. In this case the majority atoms tend to maximize the distance with the impurity, and this leads to an effectively ``attractive'' scenario, where they bunch together {(this kind of effect has also been verified in harmonically trapped systems \cite{bunchingZINNER}). We will see that this indirect attraction of the majority atoms via the impurity induces various effects in the observable properties, the most striking being the peak in low momentum for the static structure factor. Let us stress here that at least these extreme situations are representative also for ensembles of many majority atoms, since the underlying physics are similar to those explained above.

\subsection{Ansatz} Extreme cases in bosonic systems like the one depicted in Fig.~\ref{fig1} (a) allow also for exact solutions of the many-body Schr\"odinger equation that are obtained by means of Bose-Fermi mapping \cite{giraTG}, which ``maps'' the infinitely repulsive bosons to free fermions. The Jastrow-type ansatz that we employ here is based on those solutions and also on the two-body integrable Lieb-Liniger case:

\begin{equation} 
\psi_{ij}=\cos{\left[k_{ij}\left(|x_{i}-x_{j}|-1/2\right)\right]},
\end{equation}
where  $k_{ij}$ (in units of $1/L$) is related to the interaction strength $g_{ij}$ by 
$k_{ij}=2\arctan({g_{ij}/2k_{ij}})$, with $k_{ij} \in [0,\pi]$.  
The  Jastrow ansatz  for our system of three bosons reads:

\begin{eqnarray}
\psi=\mathcal{N}&\cos&{\left[k\left(|x_{i}-x_{m1}|-1/2 \right)\right]} \nonumber \\  \times &\cos&{\left[k\left(|x_{i}-x_{m2}|-1/2 \right)\right]}  \nonumber\\  \times &\cos& {\left[k'\left(|x_{m1}-x_{m2}|- 1/2 \right)\right]},
\end{eqnarray}
where $\mathcal{N}$ is a normalization constant, $x_i$ denotes the position of the impurity atom, $x_{m1}$ and $x_{m2}$ the positions of the majority atoms, and  $k,k'$ are related to $g$ and $g'$, respectively. Although here we conveniently write the ansatz for our particular system it is clear that it can be generalized for more atoms and different pairs. Indeed, for fermionic majority atoms, we modify the last term to $\sin{\left[\pi\left(x_{m1}-x_{m2}\right)\right]}$}, which respects the fermionic exchange property. Note that, contrary to the bosonic case, this term does not correspond to the exact ground state wave function for a pair of identical fermions (see \cite{manninen} for details). However, as we will show next, this expression, besides of being simpler to handle, is able to reproduce the relative correlations for the cases that we study here.

\begin{figure}
\includegraphics[scale=0.3]{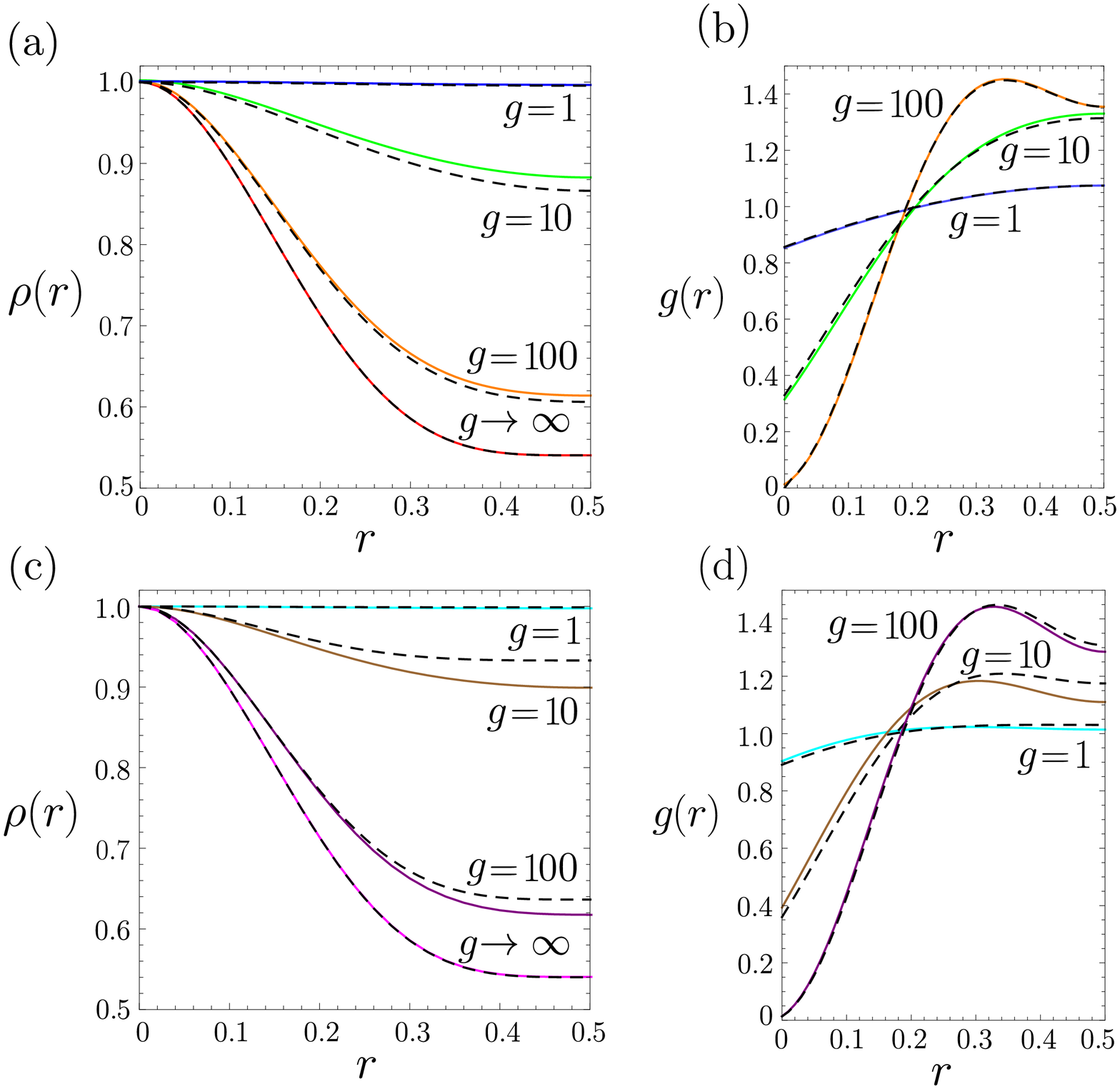} 

\caption{(Color online) Comparison of results from Jastrow ansatz and exact solutions for different interaction strengths for (a) one-body correlation function and (b) pair correlations in the integrable bosonic case and (c) one-body correlation function for the impurity fermion and (d) pair correlation for a fermionic impurity-majority pair. Black dashed lines correspond to the results obtained using the corresponding exact wave functions.}
\label{fig2}
\end{figure}
In Fig.\ref{fig2} we compare the results for correlations using the Jastrow ansatz to the exact ones given by the Bethe ansatz in the integrable cases. We particularly focus on basic quantities such as the one-body correlation function,

\begin{equation} 
\label{1bcorr}
\rho(x,x')=\int dx_{2}...dx_{N}\,\psi(x,x_2,...,x_N)\psi^{*}(x',x_2,...,x_N),
\end{equation}
and the pair correlation function

\begin{equation}\label{twobody}
g(x_1,x_2)=\int dx_{3}...dx_{N}\,|\psi(x_1,x_2,...,x_N)|^2,
\end{equation}

where we assume that the $N$-body wave functions are normalized to unity. In Fig.\ref{fig2} (a) we present the one-body correlation function with respect to the relative distance $r=x-x'$ for the bosonic case for several interaction strengths. We find that the Jastrow ansatz is in good agreement with the existing exact solutions \cite{lieb}. The same holds for the bosonic pair correlations as a function of the separation $r=x_{1}-x_{2}$, as shown in Fig.\ref{fig2} (b). As expected, in the extreme cases of vanishing and infinite repulsion (Tonks-Girardeau gas) the agreement is at best since the Jastrow ansatz coincides with the exact solution \cite{cazallila}. For the case of two majority fermions with an impurity, we compare the results of the Jastrow ansatz with the exact solution \cite{mcguire,recher} in Figs.\ref{fig2} (c) and (d). Again we find that the agreement is very good in the weak and strong interaction limits, with slight deviations for intermediate interaction strengths. Here $\rho(x,x')$ refers to the impurity fermion, and the pair correlation function is calculated for an impurity-majority pair, so $x_1=x_i$ and $x_2=x_{m1}$. For the homogeneous cases discussed here where an exact solution exists, the Jastrow ansatz provides a simple and reliable approximation of this solution. Therefore, we can state that even without a variational approach, this ansatz captures the basic physics of the systems that we discuss here. We mention that an approach on correlations for bosons in a harmonic trap has been discussed in \cite{brouzos-prl}, while correlations for the two-component trapped fermionic system have been studied in \cite{blume,lindgren}.

\section{Measurable Quantities}

\subsection{One-Body Correlation Function} 
\begin{figure*}
\includegraphics[scale=0.5]{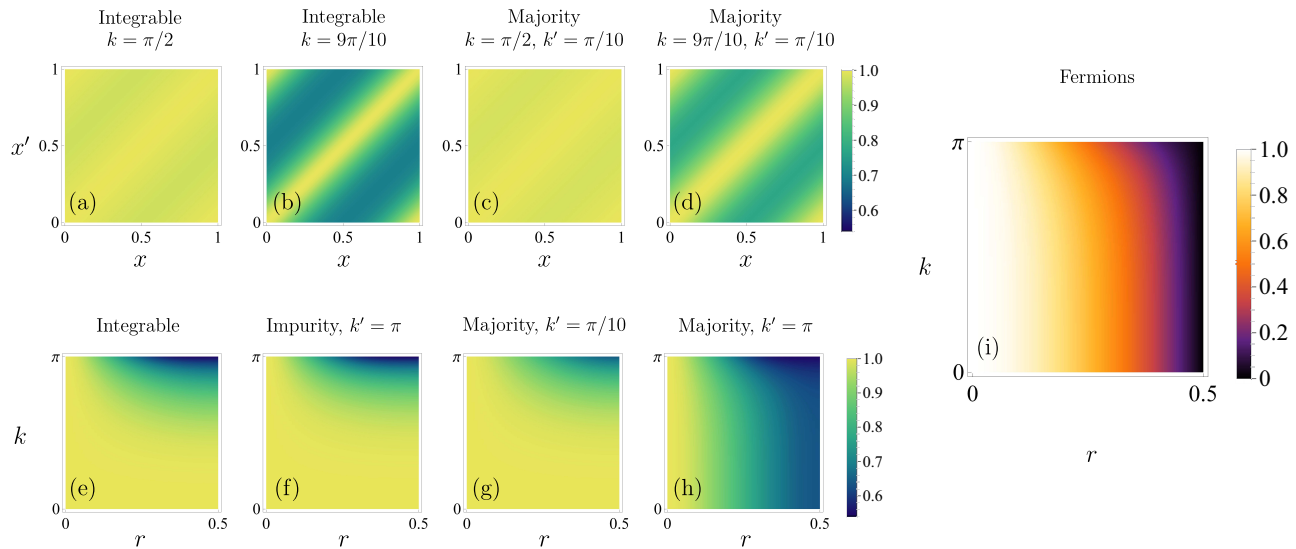} 
\caption{(Color online) One-body correlation function for representative cases interactions in the integrable, majority and impurity cases. Loss of coherence (depletion of the off-diagonal terms of $\rho(x,x')$) from (a) weak to (b) strong interaction in the integrable case, and from (c) weak to (d) strong repulsion from the impurity for the majority bosonic atoms. Behaviour of $\rho(r=|x-x'|)$ as a function of the interaction parameter $k$ in the cases of (e) integrable system, (f) impurity atom for strong majority interactions, majority atoms for (g) weak and (h) strong interactions and (i) majority fermions.}
\label{fig3}
\end{figure*}

In our system we can find either the correlation function for the impurity or the majority atoms by integrating out the coordinates of the other pair of atoms in each case. The one-body correlation function - and in particular its off-diagonal terms - characterizes the degree of coherence of a given atom of the system.
As we observe for the integrable case in Fig~\ref{fig3}.(a) when the interactions are weak or intermediate ($k=\pi/2$ corresponds to $g=\pi$) there is coherence in the ensemble since diagonal and off-diagonal terms are almost equal. 
This behaviour resembles that of the Bose-Einstein condensate, which is a coherent ensemble of many bosonic atoms. On the other hand, when the repulsive interaction becomes very strong as in Fig~\ref{fig3}.(b) ($k=9\pi/10$ corresponds to $g\approx35.7$) the atoms tend to localise and lose coherence. The diagonal $x=x'$ peak (and due to the periodic boundary conditions and symmetry the peak at distance $L$) becomes much more pronounced than the off-diagonal area around distance $L/2$. This is typical for few body one-dimensional ensembles at strong interactions which are no longer represented by a single coherent wave function but are occupying higher single-particle states (orbitals), an effect known as depletion of the condensate or fragmentation \cite{meyer,sakman,zollner,fischer}. We have a similar effect for the impurity atom when the impurity-majority interaction is strong and also for the majority atoms when the interaction between them is strong. The most interesting non-integrable case is depicted in Figs~\ref{fig3}.(c),(d) where we see the loss of coherence for the majority atoms solely due to the increase of the strength of the interaction with the impurity. Since the direct interaction term between the majority atoms is small ($k=\pi/10, g \approx 0.1$) one would expect that they remain coherent, but it is shown that if the impurity repels them  strongly, then they get localised (see also schematic Fig~\ref{fig1}.(c)). Since the one-body density matrix is always symmetric with respect to the diagonal (as we see in Figs~\ref{fig3}.(a)-(d)) we can depict the whole behaviour as a function of $r=x-x'$ (as done in Fig.\ref{fig2}) and of the interaction. 
Therefore we show in Fig~\ref{fig3}.(e)-(i) the whole behaviour of the correlation function as a function of $k$. The localization around $r=0$ again shows that the loss of coherence appears gradually for high k.
Very similar behaviour is obtained for the impurity atom as shown in Fig~\ref{fig3}.(f) due to the impurity-majority strong interaction. 
In Fig~\ref{fig3}.(g) the bosonic majority atom is shown to lose coherence at large impurity-majority interactions even if majority-majority coupling is rather low.
A comparison between the fermionic majority atoms and bosonic majority atoms with infinite repulsion ($k'=\pi$) shows that the fermionic atoms have more pronounced localisation. 
In other words,  the infinitely-repulsive bosonic atoms are resembling fermions in their non-local correlation also as a function of the repulsive strength with the impurity,
but they never get as radically localised as the majority fermions. 

\subsection{Momentum Distribution and Contact} 
An observable directly connected to the one-body correlation function is the momentum distribution:

\begin{equation}\label{mom_gira}
n(p)=\int dx\,dx' \rho(x,x')\,e^{-i p (x-x')},
\end{equation}
where $p=2\pi s$ and $s$ is an integer. We show in the Appendix that this is equivalent to calculating first the many body wave function in momentum space for one atom

\begin{equation} \label{fourier}
\phi(p,x_{2},...,x_{N})=\int dx_{1}\,e^{-i\,p x_{1}}\,\psi(x_{1},x_2,...,x_{N}), \nonumber
\end{equation}
and then integrating out the rest of the coordinates from its square modulus. 

\begin{figure}
\includegraphics[scale=0.35]{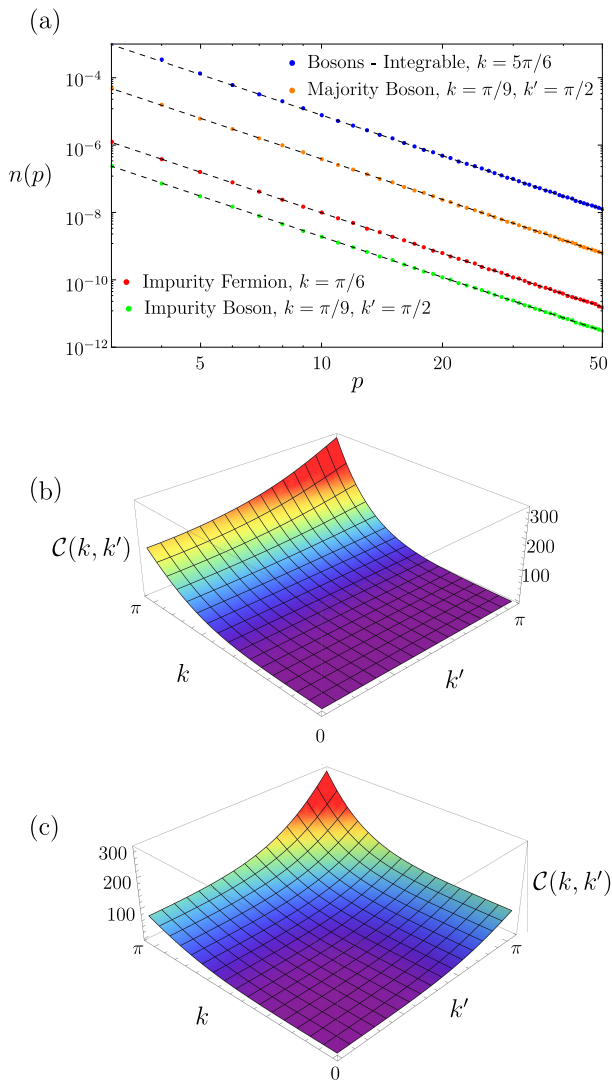}
\caption{(Color online) (a) Momentum distribution for different interactions, including integrable bosonic, majority and impurity bosons and fermionic cases ($p$ is in units of $2\pi$). Dashed black lines are the predicted values given by Eqs.~\ref{contact} (integrable) and~\ref{NI_contact} (non-integrable). (b) Contact as a function of $k$ and $k'$, obtained from Eq.~\ref{NI_contact} for the impurity atom. (c) Same as (b), but for a majority atom.}
\label{fig4}
\end{figure}

It is known that the asymptotic behaviour (the large $p$ tail) of the momentum distribution obeys an universal power law \cite{calabrese}, \cite{zwerger}:
\begin{equation}\label{tan}
n(p)=\frac{\mathcal{C}}{p^{4}},\,\, \mbox{for}\,\, p\rightarrow \infty.
\end{equation}
An expression for $\mathcal{C}$ depending on the two-body correlation function is given in \cite{olshanii}:
\begin{equation}\label{contact}
\mathcal{C}=\frac{4(N-1)\rho_{2}(0,0)}{a_{1D}^{2}},
\end{equation} 
where $a_{1D}$ is the one-dimensional scattering length, defined as $a_{1D}=-2/m g$, and $\rho_{2}(0,0)$ is the normalized two-body correlation function at vanishing distance between the atoms (an analytical expression for $\rho_{2}$ in the case of three atoms is given in \cite{yiannis-angela}, while thermodynamic limit formulas for weak and strong interactions are found in \cite{gangardt}).
The proportionality constant $\mathcal{C}$ is called \emph{contact} and is determined by the strength of the interaction. The contact characterizes all the universal properties of such systems that are independent of the details of the interaction. For a fixed interaction, it is proportional to the probability that two atoms can be found at a short distance from each other (in this case at zero separation). 
In Fig.~\ref{fig4}(a) we verify the $\mathcal{C}/p^4$ asymptotic behaviour, not only for the integrable but for all cases considered here. We also verify that Eq.~\ref{contact} is capturing the correct behaviour of the contact, for all our data points in the integrable case. For the non-integrable cases where interactions are different, there is no explicit formula in the literature for the contact. Therefore we write, in analogy to the calculations in  \cite{olshanii}, an  analytical expression for the asymptotics of momentum distribution \cite{vanja}:

\begin{eqnarray}\label{NI_contact}
n(p)\mathop{=}\limits^{|p|\rightarrow\infty}\int dx_{2}... dx_{N} \,\bigg| \sum_{j=2}^{N}(2/a_{1\text{D}}^{1j}) &&  e^{-ipx_{j}}  \nonumber \\ \times \Psi(x_{1}=x_{j},...,x_{j},...,x_{N})\bigg|^{2}\frac{1}{p^4},
\end{eqnarray}
where we identify the pre-factor as the contact. The index $j$ is inserted to account for different interactions between particle $x_{1}$ and the other particles of the ensemble. For our particular case of three particles, the interaction between the impurity and the remaining pair is the same (so $a_{1\text{D}}^{12}=a_{1\text{D}}^{13}$), and the difference from the integrable contact will be carried only in the wave function. For the contact of the majority atoms, the scattering length will be different for each pair ($a_{1\text{D}}^{21}\neq a_{1\text{D}}^{23}$).
For all non-integrable cases this formula predicts very well the  behaviour of the momentum distribution at large large momentum for our data points as is shown for exemplary cases in  Fig.~\ref{fig4}(a).
By using Eq.~\ref{NI_contact} we can then calculate the contact for all different interaction strengths, for impurity and majority atoms; the results for the contact as a function of the interaction parameters are presented in  Fig.~\ref{fig4}(b) and (c). In general the value of the contact increases with the interaction strength, but the main observation here is that the contact for the majority atoms is symmetrically dependent on $k$ and $k'$ while for the impurity the major change comes, as expected, from the increase of the interaction with the majority atoms $k$. 

\begin{figure}
\includegraphics[scale=0.35]{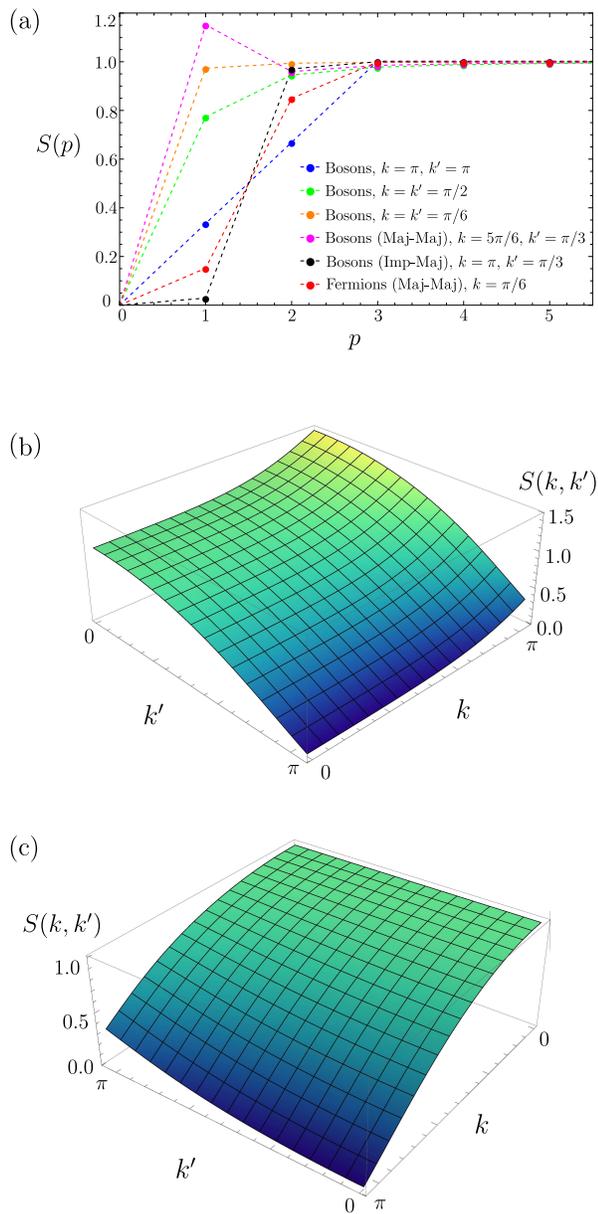} 

\caption{(Color online) (a) Low momentum region of the static structure factor for several integrable and non-integrable bosonic cases, as well the fermionic majority-majority case ($p$ is in units of $2\pi$). (b) Static structure factor as a function of $k$ and $k'$, obtained by using the majority-majority pair correlation function for bosons, with $p$ fixed as $2\pi$. (c) Same as in (b), but for an impurity-majority pair.}
\label{fig5}
\end{figure}

\subsection{Static Structure Factor}
The structure factor is a property that defines how an ensemble of atoms scatters incident radiation. Experimentally, it is usually measured by two-photon Bragg scattering, in which the atom sample is subject to two detuned laser beams. The stimulated emission of light by the atoms gives rise to interference patterns that can be measured and contain information about the structure of the sample. For different wave vectors $p$ and frequencies $\omega$ of the beams, it is possible to measure the dynamic structure factor $S(p,\omega)$ of the system. Its integral over all frequencies provides the static structure factor, defined for 1D homogeneous systems as \cite{astra}:
\begin{equation}\label{structure}
S(p)=1+\int dx_j\,dx_i\,e^{-ip(x_i-x_j)}\,[g(x_i,x_j)-1],
\end{equation}
where $p$ is again quantised as $2\pi s$ and the pair correlation $g(x_i,x_j)$ is renormalized to $N(N-1)/N^{2}$ for convenience. For our system, the pair correlation function can refer to an impurity-majority pair ($x_i=x_{i}$ and $x_j=x_{m1}$) or a majority-majority pair ($x_i=x_{m1}$ and $x_j=x_{m2}$), and $N(N-1)/N^{2}=6/9$. 

In the low momentum region and infinitely repulsive regime, hydrodynamic theory \cite{haldane} predicts that the static structure factor should behave linearly with $p$. Particularly, for the TG limit this behaviour is described by $S(p)=|p|/2\pi n$, where $n=N/L$ is the particle density. This result is characteristic of phonon excitations in many body systems, but we confirm it also in our few-body system in the integrable case as shown in Fig.~\ref{fig5}(a). For high momentum the structure factor always converges to 1 (a high speed probe would not be scattered by the cold atom ensemble).  For low momentum the behaviour of the non-integrable cases deviates much from that of the integrable, the first peak being below the TG phononic behaviour or above it.
 
Most importantly, for the majority bosons $S$ can also be over 1, an effect that indicates a quasi-crystalline structure and is not present in the integrable case. It results specifically from the fact that the repulsion with the impurity is inducing an effective attraction between the majority atoms as depicted in Fig.~\ref{fig1}(c).
This is also a characteristic of the so-called Super-Tonks gas, which is an excited state on the attractive side of the Feshbach resonance that also exhibits a peak above 1 in the structure factor at low momentum as shown in \cite{astraST}. We focus in Fig.~\ref{fig5}(b),(c) on the behaviour of the structure factor at low momentum $p=1$. In (c) we observe again this peak arising smoothly for the majority atoms and being very pronounced at low $k'$ and high $k$.  For the impurity-majority pair, $S\leq 1$, as is shown in Fig.~\ref{fig5}(c), and the main change comes from impurity-majority repulsion. 
An important issue regarding the experimental determination of the structure factor is the possibility of measuring this quantity separately for impurity-majority and majority-majority pairs. This could be achieved, for instance, by having the beams polarized, in such a way that the photons could couple differently to the spin (or pseudospin) components in the system \cite{carusotto}. Another way to obtain this difference in the coupling involves a specific detuning of the laser beams, leading to separate measures in the components of the total structure factor, as argued in \cite{combescot}.

\section{Conclusions and Outlook}

We have studied the non-local correlation functions of few-body bosonic and fermionic mixtures in a one-dimensional ring in the presence of an impurity in a different hyperfine state. The interaction strengths between impurity-majority and majority-majority pairs may differ, which renders the system non-integrable. The strength of this difference is responsible for pronounced effects in observable quantities. Most of the effects that we study on the one-body correlation function, on the contact and on the static structure factor refer to a weakly interacting majority pair which is strongly repelled by an impurity atom. In particular, this gives rise to effective attractive interactions and a pronounced peak in the structure factor. By means of our Jastrow ansatz, which is in very good agreement with the existing exact solutions for certain cases, we are able to provide results for several observable quantities in the non-integrable regime.

The experimental verification of the quantities we study is possible since recent advances in ultracold atoms techniques allowed for measurements in few-body systems. Many of our results also hold qualitatively for a larger number of majority atoms. Our approach can be extended to other cases which are relevant for experiments, like differently mixed systems and larger ensembles.

\begin{acknowledgments}
The authors thank G. E. Astrakharchick and V. Dunjko for inspiring discussions. The authors acknowledge CNPq (Conselho Nacional
de Desenvolvimento Cient{\'i}fico e Tecnol{\'o}gico) for financial
support.
\end{acknowledgments}

\appendix*
\begin{widetext}
\section{}

The momentum distribution is usually obtained by performing the Fourier transform of the one-body correlation function $\rho(x,x')$ (assuming a normalized wave function): 
\begin{equation}\label{mdist}
n(p)=\int dx\,dx' \rho(x,x')\,e^{-i p (x-x')}.
\end{equation}
We consider now $\rho(x,x')$ as given in Eq.~\ref{1bcorr}. By replacing it in Eq.~\ref{mdist}, we get

\begin{equation}\label{mom_gira}
n(p)= \int dx_{2}\,dx_{3}\,dx\,dx'\,\psi(x,x_{2},x_{3})\psi^{*}(x',x_{2},x_{3})\,e^{-i p (x-x')},
\end{equation}
which can be rewritten as
\begin{equation}\label{conta1}
n(p)=\int dx_{2}\,dx_{3}\,\int dx\,\psi(x,x_{2},x_{3})e^{-i p x}\,\int dx'\,\psi^{*}(x',x_{2},x_{3})\,e^{i p x'}.
\end{equation}

From Eq.~\ref{fourier}, $\int dx\,\psi(x,x_{2},x_{3})\,e^{-i p x}=\phi(p,x_{2},x_{3})$, so Eq.~\ref{conta1} results in
\begin{equation}\label{mdist_fin}
n(p)=\int dx_{2}\,dx_{3}\,|\phi(p,x_{2},x_{3})|^{2},
\end{equation}
which can be calculated given that an expression for $\phi(p,x_{2},x_{3})$ exists. For the momentum space representation of the wave function with respect to the coordinate $x_{1}$ in the bosonic case, this expression reads:

\begin{eqnarray}
   \sqrt{\mathcal{N}_{B}}\phi(p,x_{2},x_{3})= 
\begin{cases}
   \frac{1}{2(-4k^{2}p+p^{3})}e^{-ip(1+2x_{2}+2x_{3})}\cos{[k'(\frac{1}{2}-x_{2}+x_{3})]}\big(-i(4e^{ip(1+2x_{2}+x_{3})}k^{2} \\ -4e^{ip(1+x_{2}+2x_{3})}k^{2}-e^{2ip(x_{2}+x_{3})}(-4k^{2}+p^{2})+e^{ip(1+2x_{2}+2x_{3})}(-4k^{2}+p^{2})) \\
   \times \cos{[k(x_{2}-x_{3})]} 
   +4ie^{ip(1+x_{2}+x_{3})} (e^{ipx_{2}}-e^{ipx_{3}})k^{2}\cos{[k(1-x_{2}+x_{3})]} \\-2e^{ip(1+x_{2}+x_{3})}(e^{ipx_{2}}+e^{ipx_{3}})kp 
 (\sin{[k(x_{2}-x_{3})]}+\sin{[k(1-x_{2}+x_{3})]}) \\-ie^{2ip(x_{2}+x_{3})}p(p\cos{[k(-1+x_{2}+x_{3})]}
-2ik \sin{[k(-1+x_{2}+x_{3})]})   
   \big) , & \text{if } x_{2}>x_{3},\\
   \\

   \frac{1}{2(-4k^{2}p+p^{3})}e^{-ip(1+2x_{2}+2x_{3})}\cos{[k'(\frac{1}{2}-2x_{2}+2x_{3})]}\big(-i(4e^{ip(1+2x_{2}+x_{3})}k^{2} \\-4e^{ip(1+x_{2}+2x_{3})}k^{2}-e^{2ip(x_{2}+x_{3})}(-4k^{2}+p^{2})+e^{ip(1+2x_{2}+2x_{3})}(-4k^{2}+p^{2})) \\
   \times \cos{[k(x_{2}-x_{3})]}
   -4ie^{ip(1+x_{2}+x_{3})} (e^{ipx_{2}}-e^{ipx_{3}})k^{2}\cos{[k(1+x_{2}-x_{3})]} \\+2e^{ip(1+x_{2}+x_{3})}(e^{ipx_{2}}+e^{ipx_{3}})kp (\sin{[k(x_{2}-x_{3})]}-\sin{[k(1+x_{2}-x_{3})]}) \\-ie^{2ip(x_{2}+x_{3})}p(p\cos{[k(-1+x_{2}+x_{3})]}
-2ik \sin{[k(-1+x_{2}+x_{3})]})    \big) , & \text{if } x_{2}<x_{3},\\ 
\end{cases}
\end{eqnarray}
where $ \mathcal{N}_{B}$ is the normalization for the bosonic wave function \cite{yiannis-angela}. For the fermionic wave function, we obtain

\begin{eqnarray}
   \sqrt{\mathcal{N}_{F}}\phi(p,x_{2},x_{3})= 
\begin{cases}
   \frac{1}{2(-4k^{2}p+p^{3})}e^{-ip(1+2x_{2}+2x_{3})}\sin{[\pi(x_{2}-x_{3})]}\big(-i(4e^{ip(1+2x_{2}+x_{3})}k^{2} \\ -4e^{ip(1+x_{2}+2x_{3})}k^{2}-e^{2ip(x_{2}+x_{3})}(-4k^{2}+p^{2})+e^{ip(1+2x_{2}+2x_{3})}(-4k^{2}+p^{2})) \\
   \times \cos{[k(x_{2}-x_{3})]} 
   +2ie^{ip(1+x_{2}+x_{3})}k(2i(e^{ipx_{2}}-e^{ipx_{3}}))k\cos{[k(1-x_{2}+x_{3})]} \\
   -(e^{ipx_{2}}+e^{ipx_{3}})p(\sin{[k(x_{2}-x_{3})]}+\sin{[k(1-x_{2}+x_{3})]})-ie^{2ip(x_{2}+x_{3})} \\
   \times (-1+e^{ip})p(p\cos{[k(-1+x_{2}+x_{3})]}-2ik\sin{[k(-1+x_{2+x_{3}})]}) 
   \big) , & \text{if } x_{2}>x_{3},\\
   \\
   
  \frac{1}{2(-4k^{2}p+p^{3})}e^{-ip(1+2x_{2}+2x_{3})}\sin{[\pi(x_{2}-x_{3})]}\big(-i(-4e^{ip(1+2x_{2}+x_{3})}k^{2} \\ +4e^{ip(1+x_{2}+2x_{3})}k^{2}-e^{2ip(x_{2}+x_{3})}(-4k^{2}+p^{2})+e^{ip(1+2x_{2}+2x_{3})}(-4k^{2}+p^{2})) \\
   \times \cos{[k(x_{2}-x_{3})]} 
   -4ie^{ip(1+x_{2}+x_{3})}k^{2}\cos{[k(1+x_{2}+x_{3})]}+2e^{ip(1+x_{2}+x_{3})} \\ \times (e^{ipx_{2}}+e^{ipx_{3}}))kp
   (\sin{[k(x_{2}-x_{3})]}-\sin{[k(1+x_{2}-x_{3})]})-ie^{2ip(x_{2}+x_{3})} \\
   \times (-1+e^{ip})p(p\cos{[k(-1+x_{2}+x_{3})]}-2ik\sin{[k(-1+x_{2+x_{3}})]}) 
   \big)  & \text{if } x_{2}<x_{3}.\\
\end{cases}
\end{eqnarray}
The normalization of the wave function for the fermionic system is given by
\begin{eqnarray}
\mathcal{N}_{F}=\frac{2k^6-2k^2\pi^2-4k^4\pi^2+\pi^4+2k^2\pi^4+2k^2\pi^2\cos{(2k)}-\pi^4\cos{(2k)}+4k^5\sin{(k)}-8k^3\pi^2\sin{(k)}+4k\pi^4\sin{(k)}}{16k^2(-k+\pi)^{2}(k+\pi)^2}.\nonumber
\end{eqnarray}

\end{widetext}

\end{document}